# Title: Muscle coactivation primes the nervous system for fast and task-dependent feedback control

**Authors:** Philipp Maurus[1,2], Daniel P. Armstrong[3,4], Stephen H. Scott[3,4,5], & Tyler Cluff[1,6,7,*]

[1]Faculty of Kinesiology, University of Calgary, T2N 1N4, Calgary, Alberta, Canada

[2]Present address: National Institute of Neurological Disorders & Stroke, National Institutes of Health, Bethesda, MD 20894, USA

[3]Centre for Neuroscience Studies, Queen's University, K7L 3N6, Ontario, Canada

[4]Department of Biomedical and Molecular Sciences, Queen's University, K7L 3N6, Ontario, Canada

[5]Department of Medicine, Queen's University, K7L 3N6, Ontario, Canada

[6]Hotchkiss Brain Institute, University of Calgary, T2N 1N4, Calgary, Alberta, Canada

[7]Lead Contact

[*]Corresponding Author: tyler.cluff@ucalgary.ca

## Summary

Humans and other animals coactivate agonist and antagonist muscles in many motor actions. Increases in muscle coactivation are thought to leverage viscoelastic properties of skeletal muscles to provide resistance against limb motion. However, coactivation also emerges in scenarios where it seems paradoxical because the goal is not to resist limb motion but instead to rapidly mobilize the limb(s) or body to launch or correct movements. Here, we present a new perspective on muscle coactivation – to prime the nervous system for fast, task-dependent responses to sensory stimuli. We review distributed neural control mechanisms that may allow the healthy nervous system to leverage muscle coactivation to produce fast and flexible responses to sensory feedback.

## Muscle coactivation and the biological control of movement

Humans can perform a rich repertoire of motor skills. Further, we must adapt these actions to accommodate for changes in the environment, body, or goal during an ongoing action – to walk uphill, lift a handheld object, or complete movements with more urgency. The muscle activity that is involved in generating these actions typically follows a stereotypical reciprocal pattern. It begins with the activity of an agonist group of muscles (or 'prime movers') to initiate the movement. This activity is followed by the activation of antagonist muscle groups – muscles that act in opposite directions on the same joint – to terminate movement [1–3].

Some actions, however, involve muscle coactivation, which refers to the simultaneous activity of agonist and antagonist muscles. Muscle coactivation is often observed when humans are required to perform motor actions under challenging conditions. Muscle coactivation has also been documented when humans maintain balance in threatening environments where they may encounter disturbances while standing at heights [4,5], walk over uneven [6] or potentially slippery [7] surfaces, reach in environments that impose novel [8–10], unstable [11–13], or variable [14,15] disturbances, or manipulate handheld objects using a precision grip [16,17]. Humans also increase muscle coactivation when they perform reaching movements under more stringent time [18] and accuracy demands [19], or they receive higher reward for successful performance [20]. Increases in muscle coactivation have also been reported in other animals, such as when non-human primates use precision or power grips [21,22], reach to smaller targets that impose higher accuracy demands [cf. 23], interact with unstable mechanical environments [cf. 24], or switch from quadrupedal to bipedal gait [25]. Mice also coactivate their hindlimb muscles when they encounter mechanical perturbations while balancing on a beam [26].

Muscle coactivation has been documented extensively, but we still lack a basic understanding of why humans and other animals coactivate their muscles. Since the seminal papers by Hogan in the 1980s [27,28], many have argued that muscle coactivation allows the nervous system to alter the intrinsic viscoelastic properties of skeletal muscles by leveraging their mechanical impedance [27–31]. Mechanical impedance refers to the level of resistance of a mechanical system to a given displacement. It is influenced by the stiffness, damping, and inertial properties of the mechanical system [30]. Owing to the cellular mechanisms that power muscle contraction, muscles that are more active in isometric contractions have higher impedance and, like a spring, provide more instantaneous resistance when they are stretched from a static posture (Box 1) [32,33]. Evidence in favour of muscle stiffness and damping stems from detailed *in situ* investigation of the viscoelastic properties of isolated whole muscles in animal preparations under anaesthesia or *in vitro* examination of individual, isolated muscle fibers [34,35]. Spinal motor neurons are stimulated electrically, or the muscle preparations are bathed in calcium solutions to maintain a constant level of force under static, isometric conditions. The stiffness is measured from the onset to peak in force production [34] or within the first 20 ms [35] after the muscle preparation is stretched. Indeed, the muscle's instantaneous resistance to a disturbance increases when it is more active before it was stretched [34,35]. For a given level of contraction and velocity of displacement, muscles resist small changes in length with large changes in force. This initial or 'short-range' stiffness holds only for smaller displacements [36,37]. For large displacements beyond the muscle's elastic limit, muscles show comparatively smaller changes in force owing to inherently lower stiffness [32]. The force that a muscle can produce also depends on the velocity of contraction (force-velocity relationship), its length (force-length relationship), and whether the muscle shortens or lengthens (See [38,39]).

However, measuring the intrinsic mechanical properties of isolated muscle preparations or whole muscles under anesthesia may not provide a holistic view of muscle function under neural control. Isolated muscle preparations have the advantage of simplifying the biological system and enabling detailed investigation of the relationship between muscle activity, properties of the perturbation, and the muscle's resistance to stretch. It isolates muscle properties from neural control by examining individual muscles *ex vivo*, fibers *in vitro* or intrinsic whole muscles *in situ* under general anaesthesia. The potential downfall, however, is that they neglect contributions arising from antagonist muscles that act in the opposite direction on the same joint. Due to the force-velocity properties of skeletal muscle, there is a sudden drop in force production when an antagonist muscle is rapidly shortened. The rapid reduction in force depends on the shortening velocity for a given level of activity [39]. However, these forces are non-zero and cannot be disregarded as they would resist the action of agonist muscles that are stretched and directly counter the perturbation [37]. Muscle preparations also disregard neural control mechanisms that can alter the activation of skeletal muscles in response to rapid stretch or shortening. As a result, the link between muscle coactivation and mechanical impedance of muscles and limb or body segments has yet to be tested directly *in vivo*.

Nonetheless, the idea that muscle coactivation makes skeletal muscles resistant to stretch has had widespread impact on interpreting the control of voluntary movements in a range of biological systems *in vivo*. Altering the viscoelastic properties of skeletal muscle is thought to help circumvent delays in the neural transduction, transmission, conduction, and processing of sensory feedback [27–31] as well as delays between the onset of activity and resulting force produced by skeletal muscles [40,41]. It is widely believed that the nervous system may alter muscle coactivation, and thus, the mechanical impedance of a limb or body segment in a pre-set or feedforward manner. This concept, sometimes referred to as a 'pre-flex', is thought to stabilize the joints by providing an immediate response to stretch before the onset of muscle activity arising from delayed sensory feedback [12,42–44]. However, there are fundamental differences in the way mechanical impedance is estimated in animal studies compared to *in vivo* when humans perform motor tasks. Due to methodological limitations (see Box 1), the mechanical impedance is estimated over time windows (>120 ms post perturbation [31,45], sometimes as long as 300-500 ms [28]) that include both contributions from viscoelastic properties and neural feedback circuits [27,29,30,45–47].

It has been known for some time that muscle viscoelastic properties alone are not sufficient to counter mechanical disturbances [48–50] and require neural feedback control arising from spinal and transcortical circuits. Some have coined the term 'reflexive stiffness' [37,51–53], which lumps contributions from muscle viscoelastic properties and neural feedback responses. The problem, however, is that this approach obfuscates our understanding of the mechanical and neural benefits of muscle coactivation [cf. 46]. The approach also disregards the rich but challenging question of how the nervous system may leverage muscle coactivation to coordinate and control the responsiveness of agonist and antagonist muscles to sensory feedback. How muscle coactivation interacts with the processing of sensory feedback is an important yet comparatively unexplored problem in biological movement research.

Is altering muscle viscoelastic properties the sole purpose of muscle coactivation? There are many instances where humans and other animals use coactivation in actions and scenarios where stabilizing the limbs or body seems paradoxical. Often, these scenarios impose a need to quickly mobilize the limb(s) or body in response to sensory stimuli. Sprinters, for example, coactivate their leg muscles in the start block while waiting for the starter's gun [54]. Insects, like locusts and mantis shrimp, leverage coactivation to launch fast movements that are crucial for evading predators or capturing their next meal [55–59]. Humans and non-human primates also coactivate upper limb muscles when reaching in novel or unpredictable visual environments [60–62]. There is no obvious benefit to change muscle impedance in these contexts. In each of these examples, the task requires the participant to rapidly mobilize their body or limb(s) to correct for the imposed errors and complete movements successfully. Together, these findings suggest that muscle coactivation may allow the nervous system to expedite neural responses to sensory feedback.

Here we present a new view on muscle coactivation and its role in the neural control of sensory feedback. We focus on changes in properties of the task, body, or environment that require the nervous system to be more responsive to sensory feedback. Most of the research included in this perspective centers around the control of the upper limb in response to physical or visual disturbances while also including important details from recent findings in standing balance. We review neural circuits that process proprioceptive and visual feedback as well as those that have been implicated in generating muscle coactivation. These circuits span multiple levels of the nervous system, from structures in the spinal cord, brainstem and other subcortical areas, to the cerebral cortex. We present evidence that highlights the role of muscle coactivation in facilitating responses to sensory feedback. These rapid responses to sensory feedback may harness neural circuits throughout the entire nervous system to enable flexible, task-dependent control of voluntary motor actions. We then introduce distributed neurophysiological mechanisms that may allow the nervous system to leverage muscle coactivation to become more responsive to sensory feedback. Finally, we consider implications of muscle coactivation for prominent computational theories that have shaped our understanding of the importance of sensory feedback in biological movement control.

## Multiple neural circuits are involved in processing sensory feedback

The nervous system receives sensory information from many receptors in the body, including the eyes, muscles, tendons, joints, vestibular system, and skin when humans (and other animals) perform motor tasks [63]. These receptors provide details about the position and motion of the limb(s) and body, as well as information about the environment. Here, we focus on neural circuits involved in processing proprioceptive feedback arising from muscle spindles [52] and visual feedback from the eyes [64] during motor tasks performed with the upper limb(s). These neural feedback circuits are distributed across the spinal cord, brainstem, cerebellum, and cerebral cortex (Key Figure 1) and can alter the activation of agonist and antagonist muscles in response to sensory stimuli. The onset at which they change the activity of skeletal muscles depends on the latency at which they receive and are capable of processing sensory information (Box 2).

**Neural circuits involved in processing proprioceptive feedback**

Proprioceptive information is provided by many mechanoreceptors in our muscles, tendons, joints, and skin [65–70]. Here, we highlight the dominant role of muscle spindles in providing the nervous system with proprioceptive feedback [65]. Other reviews focus on the importance of receptors in tendons, joints, and skin that also provide the nervous system with proprioceptive feedback [65–67].

Muscle spindles are embedded in parallel with the extrafusal muscle fibers that are responsible for voluntary contractions. Spindle receptors detect changes in the amount and rate of muscle stretch following limb displacements (Box 2, for review see [52,71]). This information can rapidly alter the excitation of the agonist and antagonist muscles (Key Figure 1 & II). In the arms, muscle stretch can excite agonist muscles within ~25 ms via fast, monosynaptic connections between spindle afferents and $\alpha$ motor neurons that innervate the agonist muscles [72]. Muscle shortening can inhibit antagonist muscles within ~30 ms of the onset of a mechanical perturbation [73]. The slight delay in muscle inhibition arises from disynaptic circuits. These circuits connect spindle afferents with spinal inhibitory interneurons, which subsequently connect with $\alpha$ motor neurons innervating muscles that are antagonist to the perturbation [74,75]. These so-called short-latency responses ($SLR_{mechanical}$) are mediated entirely by spinal feedback circuits [76,77] and are thought to have simple processing capabilities at least for goal-directed actions like reaching (see [78–81] for spinal reflex complexities associated with locomotion). Their amplitude increases with the rate and amount of stretch and level of muscle activity at the onset of the perturbation [72,73,82–84]. There is some evidence, however, that the $SLR_{mechanical}$ can show flexible processing that accounts for the biomechanical properties and relative motion of individual limb segments [85,86] as well as the intended direction of goal-directed reaching movements [87].

Collaterals of muscle spindle afferents also carry proprioceptive feedback to the brainstem, cerebellum, and cerebral cortex. The information is relayed from the spinal cord to the brain by the dorsal column-medial lemniscus and spinocerebellar tracts [88]. The former is part of the shortest transcortical feedback circuit giving rise to a more flexible long-latency response ($LLR_{mechanical}$: 50 to 100 ms) compared to the $SLR_{mechanical}$. The $LLR_{mechanical}$ may alter the excitability of neurons in the spinal cord via the corticospinal, rubrospinal, and reticulospinal tracts [89]. Although the primary motor cortex may play a dominant role in implementing the $LLR_{mechanical}$ [90–94], a wide network of brain areas has also been implicated in modulating the amplitude of $LLR_{mechanical}$, including premotor and parietal areas [94,95], the cerebellum [96,97], red nucleus [98], and brainstem [99].

In contrast to the $SLR_{mechanical}$, the $LLR_{mechanical}$ of the stretched agonist muscle is a functional response that depends on features of the ongoing task, body, and environment (for review see [64,76,77,100,101]). A growing body of work highlights that the LLR$_{mechanical}$ accounts for the location of the goal target [102–106], the spatial and timing demands of the task [18,107–111], obstacles in the environment [112], interlimb coordination [113–115], limb biomechanics [116–120] and novel loads that alter these properties [121–125], and the stability of the environment [126–128] in upper limb posture control and reaching tasks.

The LLR$_{mechanical}$ of the shortened antagonists has received comparatively little attention. One reason is due to the use of background loads to control muscle activity before the onset of a mechanical perturbation (see Box 3). These small loads excite muscles that counter the load ('pre-loaded' muscles), while they evoke reciprocal inhibition of the antagonist muscles ('unloaded' muscles). Historically, most studies have looked at the stretch responses of the 'pre-loaded' muscle owing to a floor effect arising from the pre-inhibition of antagonist muscles that are shortened by the perturbation. Indeed, pre-inhibition of these muscles limits or negates their involvement in control and functional responses, as well as the extent they can be excited, owing to properties of motor neuron recruitment [cf. 129]. One exception focuses on the shortening responses of 'pre-loaded' muscles where they can act as antagonists in countering the perturbation [73]. The authors highlight a sophisticated and task-dependent inhibitory LLR$_{mechanical}$ of the pre-loaded, shortened muscle that depends on the location of the goal target and limb biomechanics.

The last ~50 years of research has provided tremendous insights into how the nervous system processes proprioceptive feedback. However, we still lack a basic understanding of how muscle coactivation alters the way proprioceptive feedback is processed. The use of background loads enhances the reciprocal pattern of muscle activity as they activate agonist muscles and inhibit antagonist muscles (for more detail see Box 3). These reciprocal patterns are also common across a range of brain areas, including the primary motor cortex, cerebellum, and brain stem [cf. 130–134]. Thus, it is difficult to generalize past findings because muscle coactivation requires fundamentally different neural control to actively and reciprocally excite groups of agonist and antagonist muscles [135]. As a result, it is unclear whether and how muscle coactivation alters the processing of even the simplest spinal feedback circuits. It also makes it unclear how the descending control and coordination of agonist and antagonist muscles help to produce long-latency responses that accommodate for features of the task, body, and environment. Further work is also required to disentangle the relative importance and interplay between muscle viscoelastic properties and neural responses to sensory feedback when fast and flexible responses to proprioceptive feedback are warranted.

**Neural circuits involved in processing visual feedback**

Receptors in the eye provide information about the position and motion of the limb(s) and features of the environment that plays an essential role in the control of voluntary limb movements and postures [136–138]. Visual perturbations, such as rapid displacements of a cursor that is projected in the participant's workspace and represents the position of their occluded finger (i.e., 'cursor jumps'; Box 2), can excite agonist muscles and inhibit antagonist muscles in the upper limbs within ~90 ms of the onset of a visual stimulus [139]. Recent evidence suggests that the earliest response, called the $SLR_{visual}$, involves a subcortical feedback pathway through the superior colliculus. The superior colliculus projects directly to the spinal cord via the tecto-spinal tract [89] and may produce express visuomotor responses [139–142]. Akin to the $SLR_{mechanical}$, the $SLR_{visual}$ possesses relatively simple, stimulus-driven processing capabilities that make it sensitive to the displacement [139,143], speed [142], and contrast [142] of the visual stimulus. Contrary to the $SLR_{mechanical}$, the $SLR_{visual}$ may also possess some degree of flexibility that has not been documented for spinal responses to proprioceptive feedback. The $SLR_{visual}$ can exploit task redundancy afforded by larger targets [23,139], depends on the predictability of the visual stimulus [144–146], and may be less sensitive to the level of background muscle activity when countering constant background loads [cf. 8,147].

Visual feedback is further processed by a transcortical circuit that includes the occipital, parietal, and frontal motor cortices [64,77,148,149]. This cortical pathway influences the activity of pools of motor neurons in the spinal cord via the corticospinal tract [89] and gives rise to a more sophisticated muscle response called the $LLR_{visual}$ (120 to 180 ms following a visual stimulus). Analogous to the $LLR_{mechanical}$, the $LLR_{visual}$ leads to rich, functional responses that account for spatial demands [23,139], obstacles [139], subject intent [cf. 141], online updates in the task goal [150], and statistical properties of the environment [151].

Akin to the $LLR_{mechanical}$, past work related to visual feedback processing has either used background loads to focus on the responses of the 'pre-loaded' muscle, examined the response of a single agonist muscle that engages to counter the visual disturbance, or disregarded the potential importance of muscle coactivation in responding to visual stimuli. It is still largely unknown how the nervous system distributes and coordinates visuomotor responses across coactivated groups of agonist and antagonist muscles.

In summary, multiple proprioceptive and visual feedback circuits are rapidly engaged following limb displacements or visual disturbances. These feedback circuits can alter the excitability of pools of spinal motor neurons that innervate the agonist and antagonist muscles. These distinct forms of disturbances recruit distinct and distributed neural circuitry that process proprioceptive or visual feedback. The common finding, however, is that rapid spinal (proprioceptive) and subcortical (visual) feedback responses tend to be relatively primitive and stimulus driven. Although transcortical feedback responses are somewhat slower, they are faster than voluntary reaction times and yield rich behavioral responses that depend on the task goal or features of the body and environment.

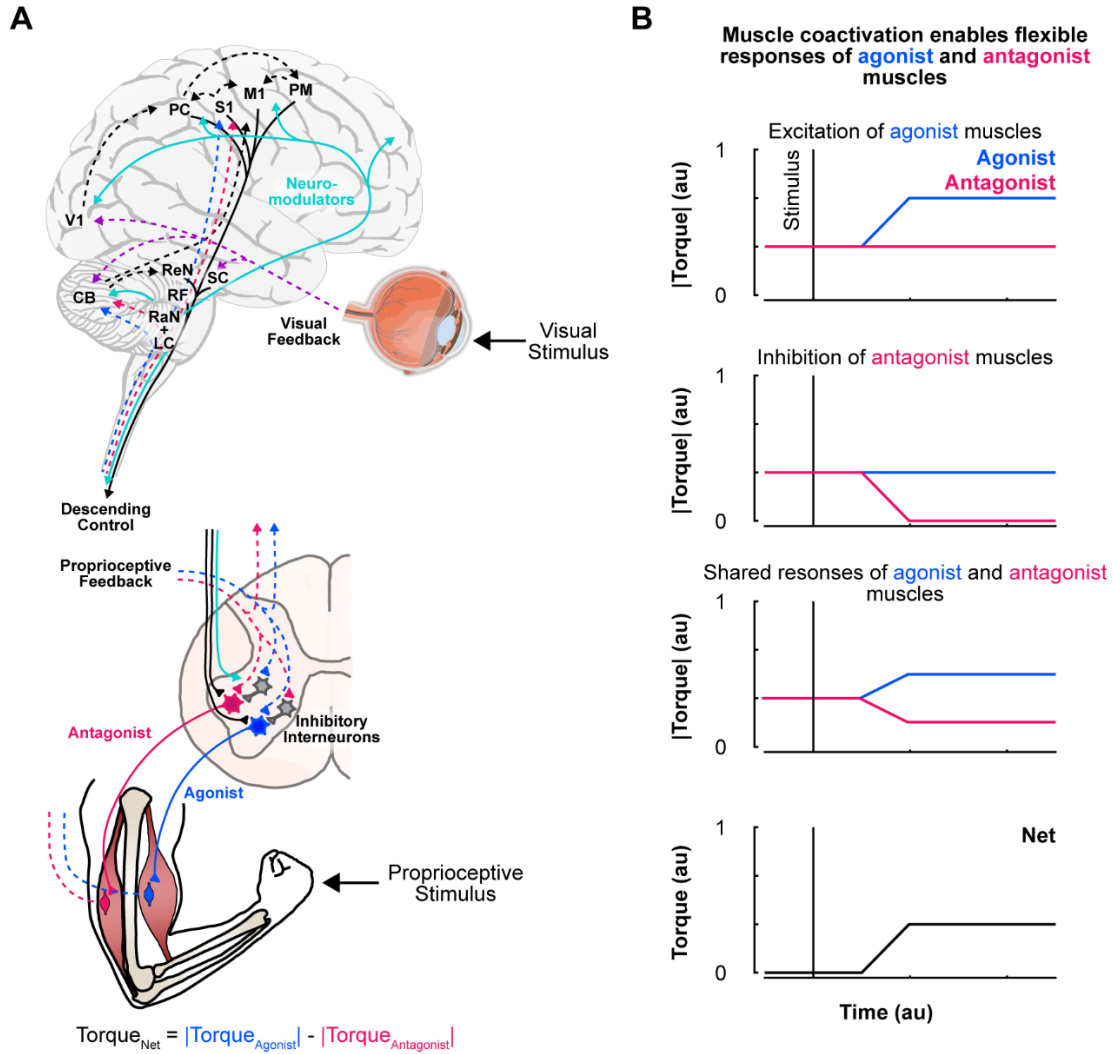

**Key Figure 1.** Feedback circuits and neural mechanisms that may prime the nervous system for fast responses to sensory feedback. **(A)** Receptors in our muscles and eyes provide sensory feedback to the nervous system. Muscle spindles detect changes in length of our muscles and send this information to the spinal cord, brainstem, cerebellum (CB), primary somatosensory cortex (S1), and primary motor cortex (M1). Visual information about the position of our body relative to the environment that we are interacting with is sent to the superior colliculus (SC) in the brainstem, cerebellum, primary visual cortex (V1), and M1. These spinal, subcortical, and cortical pathways alter the excitability of agonist and antagonist motor neurons and interneurons. Other sensory feedback pathways, such as vestibular, tactile, and auditory circuits (not shown for simplicity), can also impact the excitability of agonist and antagonist motor neurons and interneurons.
Distributed circuits, including the reticular formation (RF), red nucleus (ReN), cerebellum, parietal cortex (PC), premotor cortex (PM), and M1 have been implicated in modulating muscle coactivation of agonist and antagonist muscles within the spinal cord.
Neuromodulatory circuits, including the raphe nuclei (RaN) and locus coeruleus (LC), can alter the excitability of neurons in the spinal cord, brainstem, cerebellum, and cortex. Increasing muscle coactivation may engage these neuromodulatory circuits and may thereby alter the processing of sensory feedback.
Dashed lines indicate afferent connections. Solid lines represent efferent connections. Turquoise lines highlight projections from neuromodulatory circuits.

**(B)** The panels display the absolute torque produced by the agonist and antagonist muscles. The muscles act in opposite direction on the same joint. Note that the biceps and triceps were defined arbitrarily as the agonist and antagonist muscles, respectively. Depending on the movement direction or corrective response to a disturbance, the role of the agonist and antagonist muscles can change. Muscle coactivation enables excitation of agonist muscles, inhibition of antagonist muscles, or their shared contributions. These distributed responses can facilitate faster and greater torque production when responding to sensory feedback. Without or low levels of muscle coactivation, a floor effect prevents further inhibition of antagonist muscles, and the nervous system can only produce torque by activating agonist muscles (see also Figure III).

# Muscle coactivation primes the nervous system for fast and flexible responses to sensory feedback that are influenced by properties of the task and environment

Multiple research studies suggest that muscle coactivation may prime the nervous system for vigorous and flexible responses to sensory feedback. Muscle coactivation arises when humans and animals produce motor behaviors in more challenging environments or task demands. For example, humans increase muscle coactivation when they reach in novel, variable, or unstable mechanical environments [8,9,13–15,147,152–154] or perform movements with higher temporal urgency [1–3,18]. Under the same conditions, humans also become more responsive to proprioceptive and visual feedback [8,14,15,18,147,154]. We are only beginning to understand how coactivation and responses to sensory feedback may be linked to one another. Here, we review past work as well as emerging evidence that supports the idea that muscle coactivation may prime the nervous system for fast and flexible responses to sensory feedback.

**Muscle coactivation and responses to proprioceptive feedback**

Many studies have examined how healthy humans interact with novel, variable, or unstable mechanical environments [8,9,13–15,147,152–154]. These environments displace the arm and elicit errors that can vary between or even within a movement. Random changes in the amplitude and direction of these disturbances challenge the nervous system's ability to anticipate their consequences. Humans spontaneously coactivate their muscles when they are exposed to these environments. They also become more responsive to proprioceptive feedback, generating upregulated muscle responses that limit the displacement caused by the same mechanical perturbations [14,15,128]. Similar results have also been observed when humans perform reaching movements that impose more stringent time demands, and thus, urgency to respond to disturbances to complete movements successfully [18]. A common finding across these tasks is that the corrective responses are often shared across agonist and antagonist muscles, such that the agonists display greater excitation and the antagonists show greater inhibition in response to the same perturbations [15,18]. Paired excitation and inhibition are evident not only in the responses of fast transcortical circuits, but in the modulation of even faster spinal circuits.

Other studies have started to directly probe the causal role of muscle coactivation in modulating the nervous system's responsiveness to proprioceptive feedback. Saliba and colleagues [47] used biofeedback to modulate muscle coactivation in an upper limb posture control task. The task required healthy participants to return their arm to a fixed posture after being disturbed by brief mechanical perturbations that rapidly flexed or extended the elbow joint. Participants returned faster to the target position and displayed higher success rates when they generated higher levels of muscle coactivation before the onset of the perturbation. Muscle coactivation resulted in increased engagement of agonist and inhibition of the antagonist muscles when responding to the same perturbations. Similar results were observed by Martino and colleagues [155] when they used biofeedback to modulate muscle coactivation while healthy participants attempted to maintain standing balance following surface translations. Again, muscle coactivation engaged both agonist and antagonist muscles and enabled faster responses to the same mechanical perturbations.

Collectively, these findings support the idea that muscle coactivation may help prime the nervous system for fast and flexible responses to proprioceptive feedback. Despite the caveats of estimating stiffness *in vivo*, it is difficult to rule out contributions from intrinsic viscoelastic properties of skeletal muscle that increase stiffness and instantaneously resist the imposed limb displacements (see Box 1). Further work is needed to dissect contributions from intrinsic muscle properties and neural feedback circuits when responding to proprioceptive feedback.

**Muscle coactivation and responses to visual feedback**

There are several lines of evidence that suggest muscle coactivation may also increase the nervous system's responsiveness to visual feedback. Muscle coactivation arises spontaneously when humans and non-human primates encounter novel visuomotor rotations that do not threaten the stability of the upper limbs [60,61]. The results imply that muscle coactivation may facilitate fast responses to visual feedback since these environments require fast corrective responses to the imposed errors to complete trials successfully. Indeed, separate work has also shown that humans become more responsive to stretch following the exposure to similar novel visuomotor rotations [156]. These findings are paradoxical according to the traditional view that coactivation increases mechanical impedance and makes the limb(s) resistant to motion because visual disturbances do not impose physical limb displacements [61]. Greater mechanical impedance may even detract from vigorous responses to visual feedback by resisting rapid contraction of the agonist muscle and limb displacements that are required to correct for the imposed visual error. Thus, online corrections to visual disturbances highlight the importance of priming neural feedback circuits to alter the excitation of coactivated agonist muscles, antagonist muscles, or both muscles to mobilize the arm and produce a flexible, task-dependent response.

Recently, we systematically examined the role of muscle coactivation and responses to sensory feedback in variable visual environments [62]. Participants encountered visuomotor rotations that could vary in direction and amplitude between trials. The visuomotor rotations imposed visual errors by manipulating a cursor that represented the position of the participant's hand in their workspace. These errors required that participants quickly move their arm to correct for the visual error and complete trials successfully. We modified the difficulty of the task by increasing the amplitude of the rotations, such that participants had to correct for larger visual errors within the same amount of time to complete the trials successfully. Participants displayed higher levels of muscle coactivation when exposed to larger and thus more variable rotations. They also became more responsive to both visual and proprioceptive feedback. The corrective responses were linked to greater excitation of agonist muscles, larger inhibition of antagonist muscles, or their shared contributions that emerged in time windows associated with subcortical and transcortical feedback circuits. Importantly, the changes in amplitude of agonist and antagonist muscles prior to the onset of disturbances together with a greater $LLR_{visual}$ predicted the amplitude of corrective responses to sensory feedback. These findings highlighted a link between muscle coactivation and responses to sensory feedback.

Taken together, the results support the idea that muscle coactivation may prime the nervous system for fast and flexible responses to sensory feedback to accommodate features of the task and environment. Except for one study [18], it is largely unexplored whether the nervous system leverages muscle coactivation to upregulate responses to visual and proprioceptive feedback to accommodate differences in task goal, such as temporal or spatial urgency. We are only beginning to uncover the role of muscle coactivation for the biological control of movement.

## Descending control of spinal excitability to elicit muscle coactivation

Multiple, distributed brain areas have descending control over the excitability of motor neuron pools in the spinal cord. This top down control may enable the brain to set the stage for how sensory feedback is rapidly processed in spinal (proprioceptive), subcortical (visual), and transcortical feedback circuits (Key Figure 1) [89]. Changes in excitability can be achieved through direct, monosynaptic projections or indirect, disynaptic projections orchestrated by spinal interneurons (for review see [157]). Monosynaptic inputs are excitatory, while disynaptic inputs are inhibitory. These inputs may stem from neurons in the reticular formation [158–160], red nucleus [131,161], cerebellum [162], parietal cortex [157], premotor areas [163], or primary motor cortex [164].

Studies highlight a wide network of brain areas are engaged in tasks that require muscle coactivation. The evidence comes from neural recording and stimulation studies in non-human primates and mice as well as functional magnetic resonance imaging (fMRI) studies in humans. Neural recording studies have implicated the primary motor cortex [135,165], reticular formation [22], and cerebellum [21,166,167] when animals perform tasks that involve the coactivation of agonist and antagonist muscles. Specific examples include tasks that require non-human primates to manipulate handheld objects using precision or power grips, or mice to isometrically coactivate their forelimb muscles. Neural recordings of these tasks were often compared to motor behaviors that required reciprocal patterns of muscle activity. It is possible that the neural control of coactivation in these tasks may differ from situations where coactivation arises spontaneously in challenging environments or novel tasks.

Other evidence has shown that electrical stimulation of single neurons in the red nucleus in non-human primates can lead to the coactivation of agonist and antagonist muscles [131,161]. However, the red nucleus together with the rubrospinal tract may be less relevant for motor control in humans [cf. 168]. An fMRI study in humans also found that the BOLD signal in the premotor areas, putamen, and primary motor cortex correlated with controlled levels of muscle coactivation using biofeedback while participants maintained their wrist in a fixed position [169].

Direct measures of the neural substrates underlying flexible, task-dependent modulation of muscle coactivation in human arm movements are scarce. To the best of our knowledge, only a single study has investigated brain areas that are associated with muscle coactivation during goal-directed reaching movements [170]. In this study, muscle coactivation was not directly controlled but emerged spontaneously following the abrupt onset of a velocity-dependent force field – a well-known finding across motor adaptation studies [8–11]. The amplitude of the forces applied in this task depended on the velocity of motion and elicited movement errors by disturbing the participant's hand lateral to the goal target. Over multiple trials, the nervous system learned to compensate for the applied forces to produce accurate movements. Interestingly, the BOLD signal of the cerebellum and parietal cortex correlated with changes in muscle coactivation throughout adaptation to the novel dynamics. The findings suggest that these areas may be involved in controlling muscle coactivation, processing sensory feedback arising from the errors evoked by the applied forces, or perhaps both functions. In mice, neurophysiological recordings of cerebellar Purkinje cells reveal they are less active during faster forelimb movements [171]. Although the authors did not record muscle activity in this experiment, faster reaching movements lead to higher levels of muscle coactivation [1–3,18], implying a potential neural correlate of muscle coactivation in mice that requires further investigation.

Collectively, a wide network of brain areas, spanning the cortex, brainstem, and cerebellum, may be involved in modulating the excitability of the spinal cord to give rise to muscle coactivation. The nature of the computations and specific roles that each of these brain areas may play in generating muscle coactivation are largely unexplored. It is also unclear whether the computations performed and contributions from these distributed brain areas and circuitry depend on properties of the task (e.g., temporal or spatial urgency), body, or environment (e.g., novel, unstable, and variable disturbances).

# Neural mechanisms that may prime the nervous system for fast and task-dependent responses to sensory feedback via muscle coactivation

Muscle coactivation may leverage multiple, distributed mechanisms to prime the nervous system for flexible responses to sensory feedback that accommodate the demands of ongoing motor actions, as well as properties of the body and the environments in which they are performed. Increases in muscle coactivation may shift feedback responses from fast transcortical feedback pathways to even faster spinal (proprioceptive) and subcortical (visual) circuits [15,47,62]. It is unclear if these mechanisms act in parallel or independently from one another, and whether their relative contributions depend on the task goal or features of the limb and environment.

One potential benefit of muscle coactivation is that it may enable the nervous system to flexibly engage both agonist and antagonist muscles for voluntary control. By increasing muscle coactivation, the nervous system may be able to harness muscles with inherently different actions for rich, functional responses to sensory feedback. Indeed, coactivation may enable rapid excitation of the agonist muscles, inhibition of muscles that are antagonistic to the perturbation, or leverage their shared contributions via neural feedback loops (see Key Figure 1 B [14,15,18,47,155]). In turn, the ability to produce these flexible responses may facilitate the rapid upregulation of torques that quickly lead to more vigorous behavioral responses to sensory feedback [155]. Without coactivation, an inactive antagonist muscle cannot be further inhibited. Additionally, since faster shortening contractions produce less force for a given level of activation [39,172], inhibition of the antagonist muscles may be key in producing torque rapidly.

Changes in muscle coactivation may also leverage well-known motor unit recruitment principles by bringing larger motor units closer to or above their firing threshold [172–174]. Consequently, motor units can be (de)recruited faster and facilitate larger changes in the rate and amount of force produced in response to sensory feedback [173,174] when high gain control is required. Leveraging these motor unit recruitment principles, subtle increases in coactivation may also bring muscles closer to the steeper slope of the force recruitment curve, such that small changes in muscle activity have a larger influence on the muscle force output [cf. 172,175] and expedite the time required for a muscle to produce force or become inhibited. However, recent evidence suggests more flexibility in the recruitment of motor units, particularly when rapid force production is required [cf. 176].

Understanding how coactivation alters properties of spinal motor pools will require fundamental studies that explore how it maps onto a change in sensitivity across spinal and transcortical feedback loops. Past studies have shown that increases in tonic muscle activity in upper limb posture control automatically amplify spinal reflexes to the same perturbations [72,73,82–84]. The so-called 'automatic gain scaling' of spinal reflexes is often considered a nuisance property of spinal motor neuron recruitment because, for a given task goal, the amplitude of a reflexive response should depend only on the amount and rate of motion evoked by a perturbation. Gain scaling amplifies the response to proprioceptive feedback and limits the peak displacement of the arm [72,73,177], even when it may not be required by the task or beneficial. Moreover, under high levels of background muscle activity, motor responses to limb displacements may also result in overshooting the target and oscillations [47].

There are situations, however, where the task goal may require an increase in responsiveness to sensory feedback, such that leveraging the gain scaling properties of spinal reflexes via coactivation may help to expedite functional, task-dependent responses to sensory stimuli. This may motivate rethinking how the nervous system is able to leverage muscle coactivation, and thus potential gain scaling mechanisms to expedite flexible and goal-directed control in naturalistic motor actions. For example, higher feedback gains are required if provided less time or required to return to a smaller target when the arm is displaced by the same perturbation. Under these conditions, relying on gain scaling properties of spinal motor neuron recruitment may be advantageous since muscle coactivation may enable the nervous system to expedite the production of flexible, goal-directed corrections by fast spinal (proprioceptive) feedback circuits. In this case, the motor system relies more on a less flexible, but fast spinal response, while it may rely less on the more flexible, transcortical feedback loop [cf. 177]. It is unclear how the nervous system uses coactivation as a means to expedite flexible and task-dependent control.

Muscle coactivation may engage other, complementary neurophysiological mechanisms to make the nervous system more responsive to feedback. In addition to enabling the agonist and antagonist muscles to be engaged for rapid control and capitalizing on principles of motor unit recruitment, coactivation may also change the sensitivity of muscle spindles that monitor changes in muscle length (for review see [52]). Muscle spindles are thought to be active sensors with variable levels of sensitivity that can be altered by descending control via $\gamma$ motor neurons and $\beta$ motor neurons [52]. Increasing the sensitivity of spindles via $\alpha$-$\gamma$ coactivation or $\beta$ motor neurons, which innervate both muscle spindles and skeletal muscles [52,71,178–181], may provide a means for muscle coactivation to enhance responses to proprioceptive feedback evoked by the same mechanical disturbances. On the other hand, increased spindle excitability may be offset by other mechanisms like increased recurrent and presynaptic inhibition in a way that may depend on the level of coactivation [182–184]. It remains to be explored how these mechanisms interact with one another to enable fast and flexible responses to sensory feedback when warranted by task demands.

Arousal circuits that release neuromodulators may provide another potent mechanism to amplify responses to sensory feedback (for review see [185,186]). Arising in the brainstem, these circuits include structures like the locus coeruleus and raphe nuclei that synthesize and release norepinephrine and serotonin (Key Figure 1). Since arousal circuits have widespread impact on neural excitability throughout the entire nervous system, these circuits have the potential to modulate the gain of neural responses to sensory feedback in the spinal cord, brainstem, cerebellum, and cortex [185–188]. Past studies have shown that administration of a serotonergic agent, escitalopram, amplifies the $SLR_{mechanical}$ to the same mechanical disturbances. In contrast, administration of a serotonin antagonist, cyproheptadine, reduced the amplitude of the $SLR_{mechanical}$ to the same disturbances [189]. These findings highlight a powerful and diffuse neuromodulatory effect in controlling the gain of spinal responses to sensory feedback. Since the study only focused on spinal feedback responses elicited by a tendon tap, it raises the question of whether arousal can also help to amplify contributions from transcortical circuits when required by the goal or properties of the ongoing task.

A few studies hint at a relationship between arousal circuits and their influence of muscle coactivation during behavioral tasks. For example, when maintaining whole body posture at greater heights, participants spontaneously coactivate muscles in their lower limbs, are more responsive to stretch, and also display higher levels of electrodermal activity – a proxy for arousal and activation of the sympathetic nervous system [4]. Changes in pupil diameter, another measure of arousal, are also evident when humans reach in novel mechanical environments [190] that elicit muscle coactivation [cf. 8–11]. In mice, running speed correlates positively with pupil diameter [191] and neural firing of the raphe nuclei [192]. Since faster running speeds are associated with greater levels of muscle activity [cf. 193], the findings may imply a link between arousal and muscle coactivation. Future work is required to understand how these circuits relate to the level of muscle coactivation during behavior and whether they are modulated according to the demands of the task or features body and environment.

How does coactivation alter neural responses to sensory feedback? It has long been known that the goal of voluntary movements can shape neural firing patterns across a range of brain areas [90,91,96,194]. In non-human primates, changes in neural firing patterns of M1 and premotor areas become evident prior to movement onset [195,196], but predict features of the impending movement, like the distance, speed, and timing [197–199]. Moreover, changes in M1 firing can also predict features of rapid online corrections following a change in goal state [200]. These ideas have recently been formalized in terms of neural trajectories, dynamical systems that are impacted by the goal of an ongoing action, such that neural trajectories can vary markedly across limb movements with different speed, distance, direction, and preparatory activity [196,201]. It is clear that specific neurons or subpopulations of neurons can excite both agonist and antagonist muscles [26,131,161,165], which may lead to distinct neural state changes associated with tasks that require muscle coactivation [135]. Changes in neural states have also been observed during faster reaching movements [202] that are linked with muscle coactivation [cf. 1–3,18]. It is becoming evident that neural trajectories require recurrent connections that provide sensory feedback [203], raising the question of how the trajectories change with coactivation and what impact this has on responses to sensory stimuli when preparing for or performing voluntary actions. It is unclear if these state changes are also evident across various brain regions, such as the brainstem, cerebellum, visual cortex, somatosensory cortex and others, and if changes in muscle coactivation shift the neural population response into different parts of the state space to up- or downregulate responses to sensory feedback [cf. 204].

Taken together, the nervous system is equipped with multiple, distributed mechanisms that may enable it to harness muscle coactivation and increase its responsiveness to sensory feedback. It is unclear whether these mechanisms act in parallel or independently from one another and how they are modulated by the task goal or environment. The nervous system may leverage these mechanisms to not only upregulate responses to feedback but also to expedite feedback control from fast, transcortical pathways to even faster, spinal (proprioceptive) and subcortical (visual) circuits.

## Implications for computational theories of sensorimotor control

Humans and other animals spontaneously increase muscle coactivation during many motor tasks. In contrast, our limited understanding of muscle coactivation is highlighted by computational models that have rarely been able to explain why it emerges in biological control. Models grounded in Optimal Feedback Control (OFC) have provided tremendous insights into the importance of sensory feedback in voluntary motor actions [63,205,206]. Central to the theory is a control policy that consists of feedback gains that dictate how to generate responses to sensory feedback that represent the best way to attain the task goal [76,77]. These models have typically used simple actuators that do not represent physiological agonists and antagonists [14,95,108,112,207,208]. Although a recent study used an OFC model to predict influences of muscle coactivation on responses to limb displacements, muscle coactivation was not an emergent property but was instead set *a priori* [47]. The model did, however, support the idea that muscle coactivation can enable faster responses to sensory feedback by distributing responses across the agonist and antagonist muscles.

Other computational approaches often separate feedforward and feedback mechanisms [42,43]. These models assume that the main purpose of muscle coactivation is to alter the mechanical impedance of the limb in a feedforward manner without facilitating responses to feedback (see Box 1). For example, simulations by Van Wouwe and colleagues [42] showed that feedforward muscle coactivation emerged spontaneously to increase muscle stiffness, while reducing the overall muscle effort. However, the model predicted rather subtle changes in muscle coactivation compared to those observed in humans following the exposure to the same unstable environment [13]. Such models often focus solely on responses to proprioceptive feedback [42,209,210]. It is unclear how the model would deal with unpredictable visual environments, where muscle coactivation is an emergent property in humans and increased stiffness (or mechanical impedance) would provide little benefits.

Muscle coactivation is typically considered an expensive strategy [30,211]. However, it may be energetically efficient in some circumstances since it could help mitigate costly activation of the agonist [212,213] by leveraging the cost-free inhibition of antagonist muscles. Sharing the responses between agonist and antagonist muscles may also help reduce signal dependent noise, a widespread property of muscle force production that may relate to the recruitment properties of spinal motor neurons [214–216], where the (de)recruitment of larger units leads to larger fluctuations in force output [173,174]. Although coactivation may increase the baseline noise levels, inhibiting the antagonist muscles may help avoid large transient changes in excitation of the agonist muscles, while paired excitation and inhibition may prevent the nervous system from having to rely solely on the (de)recruitment of one specific group of muscles. Thus, distributed responses across agonist and antagonist muscles seem consistent with optimality principles that distribute the workload across coordinated pairs of muscles to exploit their redundancy in producing specific behavioral outputs [cf. 217]. Indeed, the same task-dependent response may be achieved by various combinations of excitation of agonist and inhibition of antagonist muscles (Key Figure 1).

In summary, despite the simplicity of linking muscle coactivation to stiffness, many models have not been able to predict increases in muscle coactivation. We reason that it reflects a lack in our understanding on how changes in coactivation influence neural circuits. Considering the idea that muscle coactivation may enable distributed feedback responses across agonist and antagonist muscles may provide a missing piece in our understanding of the control of movement across a range of biological systems.

**Concluding remarks**

The purpose of this perspective was to highlight current findings that have provided new insights into the role of muscle coactivation in the biological control of movement. We proposed that muscle coactivation may prime the nervous system for fast and task-dependent responses to sensory feedback by enabling excitation of agonist muscles, inhibition of antagonist muscles, or their shared contributions. Many distributed neurophysiological mechanisms may be engaged by muscle coactivation to alter the processing of sensory feedback, such as tuning the sensitivity of peripheral receptors, altering motor unit recruitment, arousal circuits, or task-dependent state changes in neural excitability. The nervous system may harness coactivation to shift feedback responses from fast transcortical to even faster spinal (proprioceptive) and subcortical (visual) circuits. In addition to altering responses to proprioceptive and visual feedback, the same physiological mechanisms linked to muscle coactivation may facilitate fast responses to tactile, vestibular, and auditory feedback to initiate and control movements. We hope this article will bring together scientists across many research fields, including neuroscience, biology, and neuromechanics, sparking exciting, new lines of investigation that begin to unravel the neural control of muscle coactivation and its proposed role in priming the nervous system to support our ability to perform and learn complex motor skills.

**Box 1: Muscle coactivation and stiffness**

Muscle coactivation may alter mechanical impedance, which reflects a limb's or joint's resistance to motion. Mechanical impedance comprises stiffness, damping, and inertia [30]. Although changes in the activation of a muscle could alter its stiffness and damping, many studies have focused on examining the link between activation and stiffness in isolated *in situ* or *in vivo* muscle preparations in cats or rats [34,35]. In these studies, single muscle fibers or whole muscles are activated, using solutions with different $Ca^{2+}$ concentrations or electrical stimulation, and then rapidly stretched using position-controlled perturbations (Figure I-A). The stiffness is measured within the first 20 ms [35] or until the peak force is reached following the stretch [34]. Increasing the activation of these preparations increases their stiffness, which is typically measured as the average force required to stretch the muscle the same fixed distance.

Position-controlled displacements have also been used to assess whether the nervous system modulates its stiffness *in vivo* when humans perform goal-directed reaching movements. A typical protocol involves small displacements of the hand (~8 mm) that are ramped up over 100 ms, held in position for 100 ms, and then ramped down over 100 ms to return the hand to the initial starting position (Figure I-B). The forces applied against the position-controller are measured during the last 60 to 80 ms of the hold period to estimate the stiffness [12,31,45,218], corresponding to 120-180 ms after the onset of the perturbation. In some instances, stiffness was estimated >300 ms after the onset of the displacement [cf. 28]. Due to vibrations [45] and because the inertia of the arm dominates the initial motion evoked by a mechanical disturbance [46], the rationale for measuring the force that is required to hold the muscle at a stretched length during this time window is to ensure stable force and position recordings. Thus, stiffness estimates in humans are at least 100 ms delayed compared to the animal work, during time windows where the relationship between activation and stiffness is much weaker [34]. Such estimates are also obfuscated by contributions from multiple neural feedback loops, including the spinal cord, brainstem, cerebellum, and cortex, that rapidly engage following limb displacements and can alter the activation of the stretched muscles [76,77]. Additionally, animal and human studies disregard the role of the opposing antagonist muscles. Although the force produced by the antagonist muscles is depressed following their shortening [219], their force is not negligible and opposes the agonist muscles unless they are inhibited [cf. 36,37,73]. Collectively, methodological differences in estimating stiffness across species make it difficult to generalize the relationship between muscle activation and stiffness from *in vitro/in situ* work in animals to *in vivo* studies in humans.

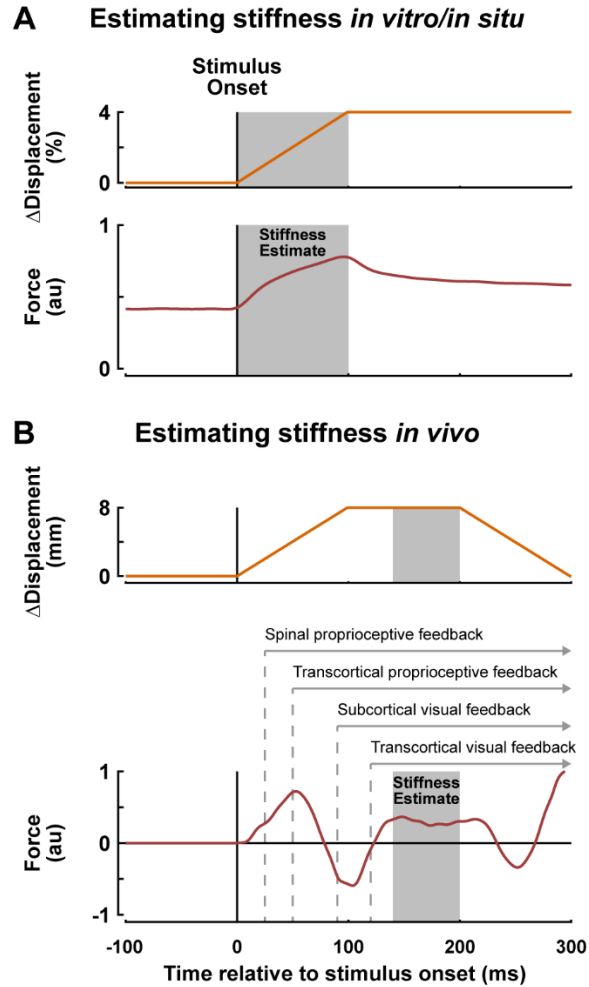

**Figure I**. Estimating stiffness *in vitro*/*in situ* in animals **(A)** and *in vivo* in humans **(B)**.
Different time windows are used to estimate the stiffness *in vitro*/*in situ* compared to *in vivo*. *In vivo* stiffness estimates in human studies are confounded by multiple proprioceptive and visual feedback circuits that rapidly engage following limb displacements.
The *in vivo* force trace was adapted from [218].

**Box 2: Probing the role of sensory feedback circuits in motor control**

Mechanical and visual disturbances have a long-standing history in probing the biological control of movement (Figure II). Robotic manipulanda [48,220–222] can apply various types of perturbations to probe the role of sensory feedback during the control of upper limb postures and reaching movements. Many studies use torque-controlled perturbations that evoke sudden limb displacements to probe the processing of proprioceptive feedback (for review see [64,76,77,100,101]). The torques are typically ramped up to peak over a short time period (~10 ms) and then held constant [76]. This approach has been used to characterize how humans and non-human primates alter their motor behavior (e.g., peak hand displacements, return times) and muscle stretch or shortening responses (i.e., $SLR_{mechanical}$ and $LLR_{mechanical}$) to accommodate various task goals or properties of the limb and environment. This technique has provided tremendous insight into how the nervous system processes proprioceptive feedback for rapid motor actions.

The robotic manipulanda are typically coupled to a virtual reality display. The display allows the presentation and precise control of visual stimuli, such as visual targets and a feedback cursor calibrated to the handle of the manipulandum or hand of the participant. The processing of visual feedback can be studied by rapidly displacing the feedback cursor [139,223] or shifting the goal target into a new location [149,150]. These stimuli have been used to characterize how humans and non-human primates alter their motor behavior (e.g., lateral velocities) [139,149] or forces applied by the participant when the stimuli were paired with error clamps (i.e., virtual force channel) that restrict hand paths to a straight line between the start position and goal target [223].

Simultaneous electromyographic recordings can reveal changes in the excitation of agonist muscles or inhibition of antagonist muscles (i.e., $SLR_{visual}$ and $LLR_{visual}$). This approach has been used to probe how the nervous system processes visual feedback for rapid motor actions.

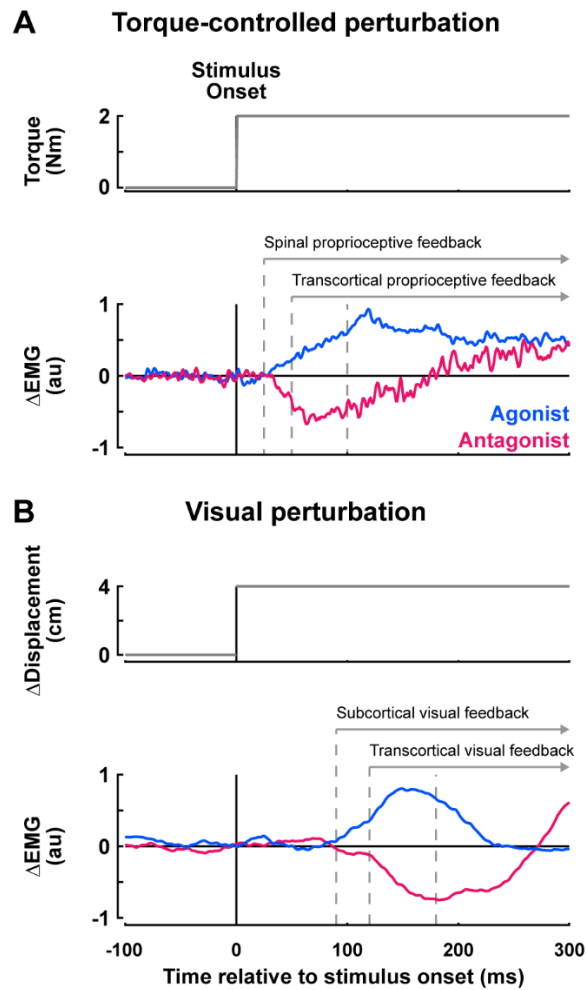

**Figure II**. Mechanical **(A)** and visual perturbations **(B)** as a probe of sensory feedback circuits in motor control.
Spinal circuits rapidly engage within ~25 ms following a mechanical perturbation, increasing the excitation of the stretched agonist muscles and inhibiting the shortened antagonist muscles (EMG). Transcortical circuits, which include the primary motor cortex, can further modulate the excitation of agonist and inhibition of antagonist muscles within ~50 ms.
Subcortical circuits via the superior colliculus are thought to alter the excitation of the agonist and inhibition of the antagonist muscles within ~90 ms following a visual perturbation. Transcortical feedback circuits, which involve the primary motor cortex, engage within ~120 ms and have more pronounced influences on the excitation of agonist and inhibition of antagonist muscles.
The EMG traces in **(A)** were adapted from [15].

**Box 3: Controlling the background muscle activity**

The use of background loads is deeply rooted in studying the processing of sensory feedback [82–84]. These loads pre-excite the motor neuron pools of the muscles of interest [76], while they inhibit the motor neurons of the antagonists to those muscles [103] (Figure III). Accordingly, they shift most of the task-dependent muscle activity to the muscle of interest. Several studies have highlighted that increasing the baseline muscle activity using background loads increases the $SLR_{mechanical}$ and to a lesser extent the $LLR_{mechanical}$ [72,73,82–84]. Due to these 'gain scaling' capabilities, researchers have historically controlled the baseline muscle activity levels as it was often deemed an inconvenience. However, the nervous system seems to possess knowledge about the gain scaling properties [177] and could thus leverage them to facilitate responses to proprioceptive feedback. Although using background loads is an attempt to control muscle activation, it is difficult to rule out sub-threshold changes in neural excitability across experimental conditions [172]. It is also unclear whether these findings generalize to muscle coactivation. Indeed, the $SLR_{mechanical}$ seems to be more pronounced when participants counter background loads prior to perturbations compared to matched levels of muscle coactivation [47]. Moreover, many studies examining H-reflexes, the electrical equivalent of muscle stretch responses, highlight reduced reciprocal inhibition as well as increased presynaptic and recurrent inhibition during tasks that require muscle coactivation compared to reciprocal inhibition of the agonist and antagonist muscles [182,183,224]. Collectively, background loads alter the spinal excitability since they require reciprocal inhibition of agonist and antagonist muscles compared to scenarios that involve muscle coactivation (cf. Figure 1 and III).

By removing the background load, it is possible to explore the richness that spontaneous levels of muscle coactivation may provide in enabling distributed responses of the agonist and antagonist muscles when responding to sensory feedback.

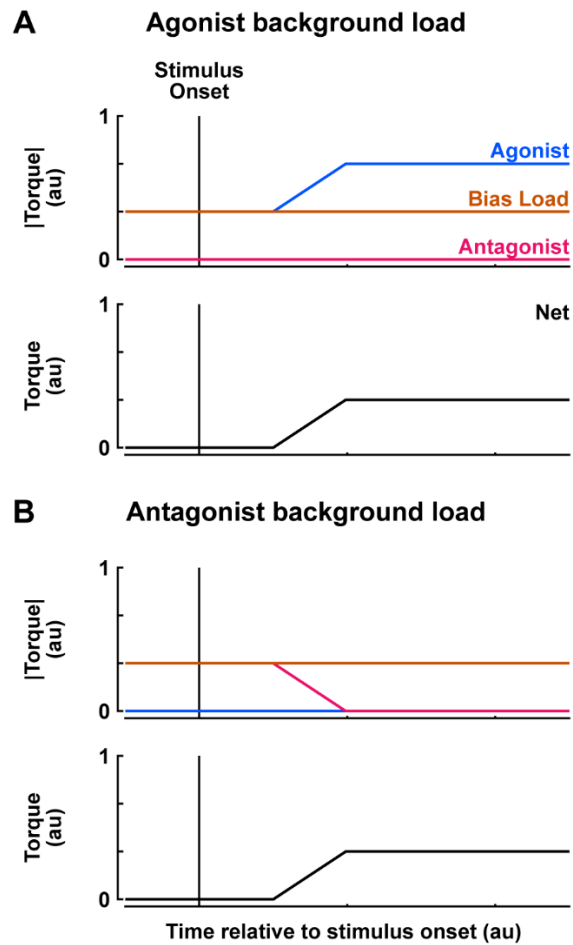

**Figure III**. Background loads alter the contributions of agonist and antagonist muscles when responding to sensory feedback.
**(A)** A background load applied to the agonist muscle shifts any response to this muscle and removes potential contributions from the antagonist muscle. Greater net torque can only be achieved by increasing contributions of the agonist muscle. **(B)** A background load applied to the antagonist muscle enables inhibition of this muscle, while muting contributions from the agonist muscle. Greater net torque can be achieved by also exciting the agonist muscle. However, background loads alter the excitability of motor neurons in the spinal cord since they require reciprocal inhibition of the agonist and antagonist muscle.

**Outstanding Questions**

- Instructed muscle coactivation enables fast responses to proprioceptive feedback. Does this finding generalize to responses to other sensory modalities (e.g., visual, vestibular, tactile, auditory)?
- Muscle coactivation often facilitates excitation of agonist and inhibition of antagonist muscles during short- and long-latency time windows. How important are the inhibitory contributions of the antagonist muscles when responding to sensory feedback? What are the relative contributions of the agonist and antagonist muscles to behavioral outcomes?
- Muscle coactivation is often evident in situations during which humans and other animals display greater levels of arousal. What is the link between muscle coactivation and the widespread impact and function of neuromodulators released by arousal circuits?
- Many physiological mechanisms may be involved in modulating responses to sensory feedback via muscle coactivation (e.g., arousal, neural state changes, muscle spindle excitability). Do these mechanisms act in parallel or independent from one another? Are their contributions dependent on the task context?
- What are the relative contributions from viscoelastic muscle properties and neural responses to mechanical perturbations?
- How does the nervous system implement muscle coactivation through a widespread network of brain areas (e.g., primary motor cortex, cerebellum, reticular formation, etc.)? What are the underlying neural computations?

- Increased muscle coactivation arises under various task constraints, such as higher temporal urgency, more stringent accuracy demands, or properties of visual and mechanical environments. How does the nervous system implement these changes in coactivation within a brain area (e.g., primary motor cortex)? Are different neural states involved depending on the behavioral goal or environment?

**Glossary**

- Muscle coactivation: the simultaneous activation of agonist and antagonist muscles – muscles that act in opposite direction at a given joint.
- Agonist muscle: muscle whose excitation initiates a voluntary movement or a corrective response to visual feedback. It also refers to the muscle that is stretched by a mechanical perturbation.
- Antagonist muscle: muscle that acts in the opposite direction on the same joint as the agonist muscle. Its excitation helps terminate a voluntary movement or a corrective response to visual feedback. It also refers to the muscle that is shortened by a mechanical perturbation.
- Mechanical impedance: refers to a resistance to a displacement. It includes stiffness (resistance to a change in position), damping (resistance to a change in velocity), and inertia (resistance to a change in acceleration). Muscle coactivation is thought to increase stiffness and damping and thus the mechanical impedance.
- $SLR_{mechanical}$ (short-latency response) & $LLR_{mechanical}$ (long-latency response): phases of electromyographic activity (EMG) elicited by a mechanical perturbation applied to a joint or limb. For the upper limbs, the $SLR_{mechanical}$ emerges between 25 to 50 ms (also referred to as R1), while the $LLR_{mechanical}$ emerges between 50 to 100 ms (also referred to as R2 and R3) after the onset of the mechanical perturbation.
- $SLR_{visual}$ & $LLR_{visual}$: phases of EMG elicited by a visual perturbation applied to a hand feedback cursor. For the upper limbs, the $SLR_{visual}$ emerges between 90 to 120 ms, while the LLR$_{visual}$ emerges between 120 to 180 ms after the onset of the visual perturbation. The SLR$_{visual}$ is also often termed express visuomotor responses or stimulus locked responses.

18. Poscente, S. *et al.* (2021) Rapid Feedback Responses Parallel the Urgency of Voluntary Reaching Movements. *Neuroscience* DOI: 10.1016/j.neuroscience.2021.07.014

19. Gribble, P.L. *et al.* (2003) Role of Cocontraction in Arm Movement Accuracy. *J. Neurophysiol.* 89, 2396–2405

20. De Comite, A. *et al.* (2022) Reward-Dependent Selection of Feedback Gains Impacts Rapid Motor Decisions. *eNeuro* 9, 1–14

21. Smith, A.M. (1981) The coactivation of antagonist muscles. *Can. J. Physiol. Pharmacol.* 59, 733–747

22. Glover, I.S. and Baker, S.N. (2022) Both Corticospinal and Reticulospinal Tracts Control Force of Contraction. *J. Neurosci.* 42, 3150–3164

23. Cross, K.P. *et al.* (2023) Proprioceptive and Visual Feedback Responses in Macaques Exploit Goal Redundancy. *J. Neurosci.* 43, 787–802

24. Addou, T. *et al.* (2015) Motor cortex single-neuron and population contributions to compensation for multiple dynamic force fields. *J. Neurophysiol.* 113, 487–508

25. Higurashi, Y. *et al.* (2019) Locomotor kinematics and EMG activity during quadrupedal versus bipedal gait in the Japanese macaque. *J. Neurophysiol.* 122, 398–412

26. Murray, A.J. *et al.* (2018) Balance Control Mediated by Vestibular Circuits Directing Limb Extension or Antagonist Muscle Co-activation. *Cell Rep.* 22, 1325–1338

—